\documentclass[column,showpacs,preprintnumbers,amsmath,amssymb]{revtex4}
\preprint{}
\usepackage{graphicx}
\usepackage{dcolumn}
\usepackage{bm}
\usepackage{mathrsfs}
\begin{document}
\title{Quantum metrology in coarsened measurement reference}
\author{Dong  Xie}
\email{xiedong@mail.ustc.edu.cn}
\affiliation{Faculty of Science, Guilin University of Aerospace Technology, Guilin, Guangxi, P.R. China.}

\author{Chunling Xu}
\affiliation{Faculty of Science, Guilin University of Aerospace Technology, Guilin, Guangxi, P.R. China.}

\author{An Min Wang}
\affiliation{Department of Modern Physics, University of Science and Technology of China, Hefei, Anhui, China.}

\begin{abstract}
We investigate the role of coarsened measurement reference, which originates from the coarsened reference time and basis, in quantum metrology. When the measurement is based on one common reference basis, the disadvantage of coarsened measurement  can be removed  by symmetry. Owing to the coarsened reference basis, the entangled state cannot perform better than the product state for a large number of probe particles in estimating  the phase. Given a finite uncertainty of the coarsened reference basis, the optimal number of probe particles is obtained. Finally, we prove that the maximally entangled state always achieves better frequency precision in the case of non-Markovian dephasing than that in the case of Markovian dephasing. The product state is more resistant to the interference of the coarsened reference time than the entangled state.
\end{abstract}

\maketitle

\section{Introduction}
With the development of quantum technology, the topic of quantum metrology, which mainly involves the estimation of physical parameters  and the improvement of measurement precision by employing quantum mechanics, has attracted considerable attraction \cite{lab1,lab2,lab3,lab4,lab5,lab6}.

Most studies have considered parameter measurement with a perfect measurement set-up\cite{lab7}. However, the primary disadvantages of this approach include the factors of noise and lossy probe particles. On the other hand, very few works have explored the estimation of physical parameters under imperfect measurement conditions. In a recent study \cite{lab8}, Fr$\ddot{o}$wis $\textsl{ et al.}$ investigated the quantum Fisher information\cite{lab9} with finite measurement precision, where the quantum Fisher information is inversely proportional to the measurement precision of the parameters. Here, we remark that coarsened measurement  includes not only the coarsened measurement precision but also the coarsened reference\cite{lab10}. A complete measurement can be divided into two steps: the first step involves setting up a measurement reference and controlling it, and the second step involves utilizing the corresponding projector to perform the final measurement (we note here that many studies address coarsened measurement in this step).
Therefore, the authors in the abovementioned study\cite{lab8} only considered the question about the investigation of quantum Fisher information in the second step. However, the coarsened reference can have a more negative function  than the coarsened measurement precision, particularly in a quantum-to-classical transition\cite{lab10}.
In other words, the coarsened measurement reference also plays a main role in quantum metrology.

Recently, $\breve{S}$afr$\acute{a}$nek et al.\cite{lab100} explored the ultimate precision limits within imperfect reference frames. However, the authors only considered a fixed definite rotation of a measurement basis. In most cases, the rotation of the measurement basis has a Gaussian distribution. Namely, the coarsened reference frames\cite{lab10} are more physically realistic than the imperfect reference frames defined in the previous study\cite{lab100}.
In this article, we investigate the role of coarsened measurement reference in quantum metrology and propose a method to reduce its adverse impact. The coarsened measurement reference originates from the coarsened measurement time and the chosen basis.
For one common reference basis, the disadvantage of coarsened measurement  can be offset by employing an even number of identical probe particles. Given a finite uncertainty of the coarsened measurement reference basis, the optimal number of probe particles is obtained in the estimating phase.
For the coarsened reference time, we find that contrary to the result in a previous study\cite{lab11,lab111}, the maximally entangled state does not achieve better precision in  the non-Markovian case than the product state when the uncertainty $\delta$ is larger than a certain value.
A previous study\cite{lab9} showed that the effect of coarsened measurement precision can be suppressed by a unitary back-squeezing operator before the final measurement. However, no unitary operator can suppress the disadvantage arising from the general coarsened measurement reference. Therefore, it is more important to control the reference accurately at first.

The rest of this article is organized as follows. In section II, we briefly introduce the measurement set-up and model. In section III, the role of the coarsened reference basis in estimating phase   is considered.
In Section IV, we discuss the coarsened measurement reference time in measuring the frequency. A concise conclusion and an outlook are included in section V.

\section{the model of coarsened measurement reference}
We consider a probe system composed of $n$ two-level particles. The Hamiltonian of each particle is given by $\hbar\omega\sigma_Z$, where $\sigma_Z$ denotes the Pauli operator with the eigenvector $(|0\rangle,|1\rangle)$.

In general, we use linear operators to perform the measurement. The form of the projective measurement operator for each particle is described with the reference basis $(|0\rangle,|1\rangle)$:
\begin{eqnarray}
&\hat{\textrm{P}}_1=(a|0\rangle+b|1\rangle)(a^*\langle0|+b^*\langle1|),\\
&\hat{\textrm{P}}_2=(b^*|0\rangle-a^*|1\rangle)(b\langle0|-a\langle1|),
\end{eqnarray}
where $|a|^2+|b|^2=1$.
When the measurement reference basis is coarsened, the reference basis $(|0\rangle,|1\rangle)$ becomes fuzzy. The above projective operator changes as follows\cite{lab10}:
\begin{eqnarray}
&\hat{\textrm{P}}_1=\int_{-\infty}^{\infty}d\theta \lambda_\Delta(\theta)U^\dagger(\theta)(a|0\rangle+b|1\rangle)(a^*\langle0|+b^*\langle1|)U(\theta),\\
&\hat{\textrm{P}}_2=\int_{-\infty}^{\infty}d\theta \lambda_\Delta(\theta)U^\dagger(\theta)(b^*|0\rangle-a^*|1\rangle)(b\langle0|-a\langle1|)U(\theta),
\end{eqnarray}
where the unitary operator $U(\theta)$ satisfies\cite{lab10}:
\begin{eqnarray}
&U(\theta)(a|0\rangle+b|1\rangle)=\cos(\theta)(a|0\rangle+b|1\rangle)+\sin(\theta)(b^*|0\rangle-a^*|1\rangle),\\
&U(\theta)(b^*|0\rangle-a^*|1\rangle)=\cos(\theta)(b^*|0\rangle-a^*|1\rangle)+\sin(\theta)(a|0\rangle+b|1\rangle).
\end{eqnarray}
$\lambda_\Delta(\theta)$ denotes the normalized Gaussian kernel
\begin{eqnarray}
\lambda_\Delta(\theta)=\frac{1}{\sqrt{2\pi}\Delta}\exp(-\frac{{\theta}^2}{2\Delta^2}),
\end{eqnarray}
where $\Delta$ represents the coarsened degree of the reference basis.

The coarsened measurement reference also includes the reference time. When measuring certain parameters such as frequency, one must choose an interrogation time. If the reference time is uncertain, the corresponding interrogation time is fuzzy.
The interrogation time $t$ will deviate from the expectation value $t_0$, with the probability
\begin{eqnarray}
p=\frac{\exp[-\frac{{(t-t_0)}^2}{2\delta^2}]}{\int_{t=0}^\infty dt\exp[-\frac{{(t-t_0)}^2}{2\delta^2}]},
\end{eqnarray}
where the range of time is $0\leq t<\infty$.

The famous Cram$\acute{e}$r--Rao bound\cite{lab12} offers a very good parameter estimation:
\begin{eqnarray}
(\delta x)^2\geq\frac{1}{N\mathcal{F}[x]},
\end{eqnarray}
where $N=T/t$ represents total number of experiments given by the fixed total time $T$, and $t$ the interrogation time. $\mathcal{F}(x)$ denotes the Fisher information, which is defined as
\begin{equation}
\mathcal{F}(x)=\sum_k P_k(x)[\frac{d\ln[P_k(x)]}{dx}]^2,
\end{equation}
where $P_k(x)$ denotes the probability of obtaining the set of experimental results $k$ for the parameter value $x$.
The coarsened measurement reference will reduce the amount of the Fisher information, leading to a reduction in the precision of the parameter.
\section{Measuring phase in coarsened reference basis}
Here, we consider the measurement of the phase of the probe system. The final measurement precision depends on the initial state. In a perfect reference basis, the initial maximally entangled state $|0\rangle^{\otimes n}+|1\rangle^{\otimes n}$ can aid in enhancing the resolution of phase $\phi$ to the Heisenberg limit: $\delta \phi\propto\frac{1}{n}$. However, the product state $(|0\rangle+|1\rangle)^{\otimes n}$ only achieves the quantum limit.

\subsection{Common coarsened reference}
In the coarsened reference basis, the final precision is influenced by the uncertain reference basis.
When the initial state of the probe system is the maximally entangled state, a phase $\phi$ is encoded after a certain time: $|0\rangle^{\otimes n}+\exp(in\phi)|1\rangle^{\otimes n}$. When the local
generator of the phase change is $\sigma_Z$, then the optimal
measurement operator is $\sigma_X$. Therefore, the optimal linear projector for each particle in the coarsened reference basis can be written as
\begin{eqnarray}
\hat{\textrm{P}}_1=1/2\int_{-\infty}^{\infty}d\theta \lambda_\Delta(\theta)U^\dagger(\theta)(|0\rangle+|1\rangle)(\langle0|+\langle1|)U(\theta),\\
\hat{\textrm{P}}_2=1/2\int_{-\infty}^{\infty}d\theta \lambda_\Delta(\theta)U^\dagger(\theta)(|0\rangle-|1\rangle)(\langle0|-\langle1|)U(\theta),
\end{eqnarray}
where the unitary operator can be chosen as $U(\theta)=\exp(-i\theta\sigma_Z)$.

We need to determine whether there is a common origin leading to the $n$ coarsened reference bases, for example, when we choose the measurement operator to be $\sigma_X$, i.e., when we measure the system along the X direction. We consider n measurements along the X direction based on the common coordinate system. When the common coordinate system is coarsened, the n measurement operators are coarsened synchronously. In other words, the $n$ linear measurement operators in the common coordinate system are correlated.
Thus, with synchronously coarsened reference bases, the corresponding projector becomes
\begin{eqnarray}
\hat{\textrm{P}}(\eta_1,\eta_2,...,\eta_n)&=&\int_{-\infty}^{\infty}d\theta \lambda_\Delta(\theta)[\frac{1}{2}U^\dagger(\theta)(|0\rangle+(-1)^{\eta_1}|1\rangle)(\langle0|+(-1)^{\eta_1}\langle1|)U(\theta)]\nonumber\\
&\otimes& [\frac{1}{2}U^\dagger(\theta)(|0\rangle+(-1)^{\eta_2}|1\rangle)(\langle0|+(-1)^{\eta_2}\langle1|)U(\theta)]\nonumber\\
&...&[\frac{1}{2}U^\dagger(\theta)(|0\rangle+(-1)^{\eta_n}|1\rangle)(\langle0|+(-1)^{\eta_n}\langle1|)U(\theta)],
\end{eqnarray}
with $\eta_1,\eta_2,...\eta_n=0,1.$
We can subsequently obtain the probability:
\begin{eqnarray}
&\textrm{P}_1(\eta_1,\eta_2,...,\eta_n)=\frac{1+\exp(-n^2{\Delta}^2)}{n}\cos^2(n\phi/2)+\frac{1-\exp(-n^2{\Delta}^2)}{n}\sin^2(n\phi/2), \ \textmd{for}\  (-1)^{(\eta_1+\eta_2+...+\eta_n)}=1;\\
&\textrm{P}_2(\eta_1,\eta_2,...,\eta_n)=\frac{1+\exp(-n^2{\Delta}^2)}{n}\sin^2(n\phi/2)+\frac{1-\exp(-n^2{\Delta}^2)}{n}\cos^2(n\phi/2), \ \textmd{for}\  (-1)^{(\eta_1+\eta_2+...+\eta_n)}=-1.
\end{eqnarray}
Substituting the above probabilities into Eq.(10), the Fisher information is obtained as
\begin{equation}
\mathcal{F}(\phi)=\frac{n^2\sin^2(n\phi)\exp(-2n^2{\Delta}^2)}{1-\cos^2(n\phi)\exp(-2n^2{\Delta}^2)}.
\end{equation}
When $n\phi=k\pi/2$ with odd $k$, the resolution of the phase  is given by
\begin{equation}
\delta\phi|_e=\frac{1}{\sqrt{Nn^2\exp(-2n^2{\Delta}^2)}}
\end{equation}
Subsequently, we can obtain the optimal resolution for $n=\frac{1}{\sqrt{2}\Delta}$,
\begin{equation}
\delta\phi|_e=\frac{\sqrt{2\Delta^2 e}}{\sqrt{N}}.
\end{equation}

It is easy to obtain the optimal resolution of the phase in the coarsened reference basis with the initial probe in the product state.
\begin{equation}
\delta\phi|_p=\frac{1}{\sqrt{Nn\exp(-2{\Delta}^2)}}.
\end{equation}
For $n\gg1$, $\delta\phi|_e\propto\sqrt{e^{n^2}}\gg\sqrt{\frac{1}{n}}$.
Obviously, when n measurement operators are coarsened synchronously, the entangled state will not perform better than the product state for large values of n.

In order to reduce the influence of the coarsened reference basis, we use a unitary operator to transform the encoded state $|0\rangle^{\otimes n}+\exp(in\phi)|1\rangle^{\otimes n}$ into $|0101,...\rangle+\exp(in\phi)|1010,...\rangle$. When the number of particles is even, the effect of the coarsened measurement reference is completely eliminated, and the disadvantageous factor is negated completely. This can be verified by utilizing the projection operators Eq.(13) to measure $|0101,...\rangle+\exp(in\phi)|1010,...\rangle$. The corresponding probability is given by
\begin{eqnarray}
&\textrm{P}_1(\eta_1,\eta_2,...,\eta_n)=\frac{2}{n}\cos^2(n\phi/2), \ \textmd{for}\  (-1)^{(\eta_1+\eta_2+...+\eta_n)}=1;\\
&\textrm{P}_2(\eta_1,\eta_2,...,\eta_n)=\frac{2}{n}\sin^2(n\phi/2), \ \textmd{for}\  (-1)^{(\eta_1+\eta_2+...+\eta_n)}=-1.
\end{eqnarray}
 Substituting the above equations into Eq.(10), we recover the Heisenberg limit,
\begin{eqnarray}
\delta\phi=\frac{1}{\sqrt{Nn^2}}.
\end{eqnarray}
For odd particles, the final resolution of the phase is $\delta\phi=\frac{1}{\sqrt{Nn^2\exp(-2{\Delta}^2)}}$.
Obviously, the use of a unitary transformation before measurement to obtain an appropriate state can improve the measurement precision in the common measurement reference.
\subsection{Independent coarsened references}
In general, the n linear measurement operators are coarsened independently. Without loss of generality, we suppose that the coarsened degrees of n coarsened reference bases are the same. Thus, the corresponding measurement projectors for n particles are given by
\begin{eqnarray}
\hat{\textrm{P}}(\eta_1,\eta_2,...,\eta_n)=\int_{-\infty}^{\infty}d\theta_1 \lambda_\Delta(\theta_1)[\frac{1}{2}U^\dagger(\theta)(|0\rangle+(-1)^{\eta_1}|1\rangle)(\langle0|+(-1)^{\eta_1}\langle1|)U(\theta_1)]\nonumber\\
\otimes \int_{-\infty}^{\infty}d\theta_2 \lambda_\Delta(\theta_2)[\frac{1}{2}U^\dagger(\theta_2)(|0\rangle+(-1)^{\eta_2}|1\rangle)(\langle0|+(-1)^{\eta_2}\langle1|)U(\theta_2)]\nonumber\\
...\int_{-\infty}^{\infty}d\theta_n \lambda_\Delta(\theta_n)[\frac{1}{2}U^\dagger(\theta_n)(|0\rangle+(-1)^{\eta_n}|1\rangle)(\langle0|+(-1)^{\eta_n}\langle1|)U(\theta_n)],
\end{eqnarray}
with $\eta_1,\eta_2,...\eta_n=0,1.$

For the initial probe particles in the maximally entangled state $|0\rangle^{\otimes n}+\exp(in\phi)|1\rangle^{\otimes n}$, we can obtain the probability distribution:
\begin{eqnarray}
&\textrm{P}_1(\eta_1,\eta_2,...,\eta_n)=&\frac{1+\exp(-n{\Delta}^2)}{n}\cos^2(n\phi/2)+\frac{1-\exp(-n{\Delta}^2)}{n}\sin^2(n\phi/2), \textmd{for}\  (-1)^{(\eta_1+\eta_2+...+\eta_n)}=1;\\
&\textrm{P}_2(\eta_1,\eta_2,...,\eta_n)=&\frac{1+\exp(-n{\Delta}^2)}{n}\sin^2(n\phi/2)+\frac{1-\exp(-n{\Delta}^2)}{n}\cos^2(n\phi/2), \textmd{for}\  (-1)^{(\eta_1+\eta_2+...+\eta_n)}=-1.
\end{eqnarray}

Using the same method described in the previous subsection, we can obtain the optimal resolution of phase
\begin{equation}
\delta\phi=\frac{1}{\sqrt{Nn^2\exp(-2n{\Delta}^2)}}.
\end{equation}
In comparison with the resolution in Eq.(17), the resolution in the independent coarsened reference is better than that in the common coarsened reference. However, the influence of the independent coarsened reference cannot be offset by the method described in the previous subsection because the coarsened references are not synchronous.
The maximally entangled state does not always perform better than the product state for large n. The optimal resolution for the entangled state in the independent coarsened references is obtained with $n=\frac{1}{\Delta^2}$.
\section{Measuring frequency in coarsened measurement time reference}
In this section, we consider measurement of the frequency $\omega$ of the probe system. Further, each probe particle is subjected to a system--environment interaction that induces pure dephasing. The interaction Hamiltonian is of the form $\sigma_Z\otimes B$, where $B$ represents some operator of the environment. The reduced density matrix of each probe particle satisfies\cite{lab111}
\begin{eqnarray}
\rho_{jj}(t)&=&\rho_{jj}(0),\\
\rho_{01}(t)&=&\rho_{01}(0)e^{-2\gamma(t)},
\end{eqnarray}
with $j=0,1$.

When the initial state of the probe system is the product state $(|0\rangle+|1\rangle)^{\otimes n}$, we can use the Ramsey spectroscopy set-up\cite{lab13} to measure the frequency in the coarsened reference time. That is, substituting the projection operators Eqs.(3-4) and the probability Eq.(8) into Eq.(10) according to the method reported in\cite{lab111},
we can obtain the resulting single particle signal
\begin{eqnarray}
\textrm{P}_0=\frac{1}{{\int_{t=0}^\infty dt\exp[-\frac{{(t-t_0)}^2}{2\delta^2}]}}\int_{t=0}^{\infty}dt\exp[-\frac{{(t-t_0)}^2}{2\delta^2}]\frac{1+\cos(\phi t)\exp(-\gamma(t))}{2},
\end{eqnarray}
where $t_0$ denotes the optimal interrogation time without the coarsened reference, and $\phi$ the detuning between the frequency of the external oscillator and that of the probe particle $\omega$ \cite{lab11}.

Subsequently, via calculation of the Fisher information, the uncertainty of the frequency is obtained in the coarsened reference time as
\begin{eqnarray}
\delta\omega^2=\frac{[1-\langle\cos(\phi t)\exp(-\gamma(t))\rangle^2]\langle t\rangle}{n T\langle t\sin(\phi t)\exp(-\gamma(t))\rangle^2},
\end{eqnarray}
where $\langle f(t)\rangle=\int_{t=0}^{\infty}dt\exp[-\frac{{(t-t_0)}^2}{2\delta^2}]f(t)/{\int_{t=0}^\infty dt\exp[-\frac{{(t-t_0)}^2}{2\delta^2}]}$.
The value of $t_0$ is given by
\begin{eqnarray}
2t\frac{d\gamma(t)}{dt}|_{t=t_0}=1
\end{eqnarray}
The value of $\phi$ is given by $\phi t_0=k \pi/2$ with  odd $k$. For reducing the influence of the coarsened reference time, the value of $k$ should be 1.
\begin{figure}[h]
\includegraphics[scale=1]{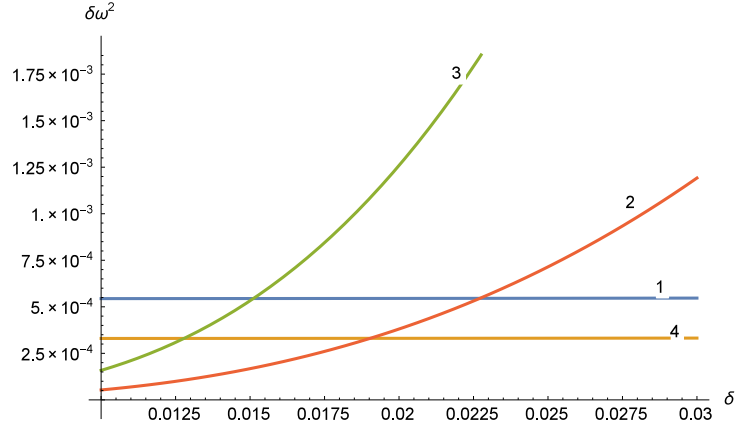}
 \caption{\label{fig.1}Precision of frequency $\delta \omega^2$ as a function of the uncertainty of reference time $\delta$.  Curve 1 represents the precision in the case of Markovian dephasing with the initial product state. Curve 2 represents the case of non-Markovian dephasing with the maximally entangled state. Curve 3 represents $10^{-8}$ times the precision in the case of Markovian dephasing with the maximally entangled state, where the scaling factor of $10^{-8}$ is used to plot four curves in one diagram because the precision in this case increases considerably faster than in other cases with the given parameters for the current scope of $\delta$. Curve 4 represents the precision in the case of non-Markovian dephasing with the initial product state. When the uncertainty $\delta$ is larger than a certain value, curve 2 (curve 3)  exhibits higher precision than  curve 4 (curve 1). The following parameters are used here: $n=10^4$, $\gamma (0)=1$, $T=1$.  }
 \end{figure}

A similar calculation can be performed for the initial state of the probe system in the maximally entangled state $|0\rangle^{\otimes n}+|1\rangle^{\otimes n}$. The corresponding resolution of the frequency is obtained as
\begin{eqnarray}
\delta\omega^2=\frac{[1-\langle\cos(n\phi t)\exp(-n\gamma(t))\rangle^2]\langle t\rangle}{n^2 T\langle t\sin(n\phi t)\exp(-n\gamma(t))\rangle^2},
\end{eqnarray}
In this case, the value of optimal interrogation time $t_0|_e$ is given by
\begin{eqnarray}
2nt\frac{d\gamma(t)}{dt}|_{t=t_0|_e}=1.
\end{eqnarray}
The corresponding value of $\phi$ is given by $n\phi t_0|_e= \pi /2$.

 From a previous study\cite{lab11,lab111}, it is known that with perfect measurement reference time, the product and the maximally entangled preparations of the probe system achieve the same resolution of frequency when subject to Markovian dephasing $\gamma(t)=\gamma(0)t$. Further, when subject to general non-Markovian dephasing $\gamma(t)=\gamma(0)t^2$, the maximally entangled state can perform better than the product state, thereby leading to resolution beyond the quantum limit.  Here, non-Markovian dephasing generally occurs when condensed matter systems are subjected to non-Markovian environments with characterized long correlation times and/or structured spectral features\cite{lab14,lab15,lab16,lab17,lab18}. The universal time dependence $\gamma(t)=\gamma(0)t^2$ is the fundamental basis of the quantum Zeno effect\cite{lab19,lab20}

 From the above equations, we note that the resolution must decrease because the coarsened reference time makes it impossible to perform measurements at the optimal interrogation time. As shown in Fig.1, the maximally entangled state does not achieve better precision than the product state when the uncertainty $\delta$ is larger than a certain value. Namely, the product state is more resistant to the interference of coarsened reference time than the entangled state. Further, the maximally entangled state always achieves better precision in the case of non-Markovian dephasing than that in the case of Markovian dephasing.

\section{conclusion and outlook}
In this study, we investigated the role of coarsened measurement reference basis and time in quantum metrology.
When subject to synchronously coarsened reference bases, a unitary transformation before measurement can aid in eliminating the disadvantages arising from coarsened measurement conditions. In the independent coarsened basis reference, the maximally entangled state cannot overcome the standard quantum limit in measuring the phase. Further, the maximally entangled state does not always perform better than the product state for a large number of particles n.
In the coarsened time reference, the product state is more resistant to the interference of coarsened reference time than the entangled state. The maximally entangled state always achieves better precision in the case of non-Markovian dephasing than in the case of Markovian dephasing.

The coarsened reference can exert a more significant influence in quantum metrology than the coarsened measurement precision. Namely, in typical noisy metrological scenarios (depolarization,
dephasing, and particle loss), the Fisher information scales linearly with $n$, as opposed to the Heisenberg scaling $n^2$. Here, in the scenario of a noisy reference, the Fisher information tends to zero exponentially in the system size  (Eq. 16) for any fixed finite $\Delta$. It is necessary to reduce the the negative effects from the coarsened reference, which is very important in experiments.
\section*{Acknowledgement }
This work was supported by 2016GXNSFBA380227 and the National Natural Science Foundation of China under Grant No. 11375168.

\end{document}